\documentclass[aps,pra,epsf,superscriptaddress,amsmath,amssymb,amsfonts,twocolumn,showpacs,nofootinbib]{revtex4-1}

\usepackage{amssymb}
\usepackage{amsmath}
\usepackage{amssymb}
\usepackage{subfigure}
\usepackage{color}
\usepackage{mathrsfs} 
\usepackage{float}
\usepackage{comment}
\usepackage{soul}

\usepackage{hyperref}

\usepackage{epsfig}
\usepackage{dcolumn}
\usepackage{bm}
\usepackage{braket}
\usepackage{mathtools}
\usepackage{graphicx,color,xcolor}
\usepackage{hyperref}
\newcommand{\abs}[1]{\left| #1 \right|} 
\usepackage{multirow}

\usepackage[normalem]{ulem} 

\begin{document}

\pagenumbering{roman}

\setcounter{page}{0}


\setcounter{page}{0}    
\pagenumbering{arabic}

\title{Two-dimensional solitons 
in extended GPE models with Lee-Huang-Yang corrections} 

\author{G. N. Koutsokostas}
\affiliation{Department of Physics, National and Kapodistrian University of Athens,
Panepistimiopolis, Zografos, Athens 15784, Greece}
\author{F. Bristy}
\affiliation{Department of Physics and LAMOR, Missouri University of Science and Technology, Rolla, MO 65409, USA}
\author{E.-C. Psychogiou}
\affiliation{Department of Physics, National and Kapodistrian University of Athens,
Panepistimiopolis, Zografos, Athens 15784, Greece}
\author{G. A. Bougas}
\affiliation{Department of Physics and LAMOR, Missouri University of Science and Technology, Rolla, MO 65409, USA}
\author{G. C. Katsimiga}
\affiliation{Department of Physics and LAMOR, Missouri University of Science and Technology, Rolla, MO 65409, USA}
\author{S. I. Mistakidis}
\affiliation{Department of Physics and LAMOR, Missouri University of Science and Technology, Rolla, MO 65409, USA}
\author{P.~G. Kevrekidis}
\affiliation{Department of Mathematics and Statistics, University of Massachusetts,Amherst, MA 01003-4515, USA}
\author{D. J. Frantzeskakis}
\affiliation{Department of Physics, National and Kapodistrian University of Athens,
Panepistimiopolis, Zografos, Athens 15784, Greece}


\begin{abstract} 

We investigate the existence and dynamics of two-dimensional solitary waves in a quantum droplet environment 
described by the extended Gross-Pitaevskii equation featuring logarithmic mean-field and Lee-Huang-Yang interactions. 
In the modulationally stable regime of the background, we employ suitable multiscale asymptotic methods to derive effective nonlinear integrable models corresponding to the 
Kadomtsev-Petviashvili and Davey-Stewartson equations. 
Based on these reduced models, we 
construct approximate analytical solutions describing line solitons, algebraically localized lump solitons, ring solitons, and exponentially localized dromions embedded on the droplet background. 
The dynamical robustness of these solutions is monitored through numerical simulations. 
Line, lump and ring solitons stay closest to the theoretical predictions, although progressively deviate due to the emergence of small-amplitude radiation, 
while dromions depart from their analytical waveform the most, although they roughly maintain their shape. 
Our results unveil unprecedented  multidimensional soliton solutions in models featuring the competition of mean-field and
quantum fluctuations and as such are amenable to current ultracold atom experiments.

\end{abstract}

\maketitle

\section{Introduction} 

Ultracold atoms offer versatile platforms for the exploration of quantum many-body phases~\cite{Bloch_many} and associated nonlinear wave phenomena~\cite{kevrekidis2008emergent}. 
This is due to their exquisite tunability in terms of the involved system parameters ranging from dimensionality~\cite{bloch2012quantum}, external confinement~\cite{wolswijk2025trapping,navon2021quantum} and interatomic interactions whose sign and strength are controlled via Fano-Feshbach resonances~\cite{chin2010feshbach,kohler2006production}. 
In particular, the regime of attractive interactions is traditionally challenging to assess owing to the unstable character of the many-body system~\cite{Dalfovo_review,donley2001dynamics} accommodating intriguing bound states of matter. These include molecular gases~\cite{langen2024quantum,schindewolf2025few}, Efimov states~\cite{ferlaino2010forty,naidon2017efimov}, and quantum droplets~\cite{petrov2015quantum,luo2021new}, but also self-localized nonlinear waves such as the two-dimensional (2D) Townes soliton~\cite{Chen_Townes,Bakkali_Townes} and the one-dimensional (1D)  Peregrine soliton ~\cite{Peregrine_exp,bougas2025observation}. 

Quantum droplets are incompressible liquid-type states~\cite{luo2021new,mistakidis2023few} 
---proposed in the work of~\cite{petrov2015quantum,petrov2016ultradilute}---
that are sustained  due to the presence of quantum fluctuations 
and their competition with mean-field (effectively nonlinear) effects.
They are often described by the dimension dependent~\cite{Zin_quantum_2018,pelayo2025phases} first-order Lee-Huang-Yang (LHY)~\cite{lee1957eigenvalues} correction to the mean-field energy functional --see also Refs.~\cite{parisi2019liquid,mistakidis2021formation,ota2020beyond,Hu_beyond_LHY} for beyond LHY theory studies. 
They have been originally observed in cold dipolar gases~\cite{chomaz2022dipolar} and afterwards in both homonuclear~\cite{cabrera2018quantum,cheiney2018bright,semeghini2018self,ferioli2019collisions} and heteronuclear~\cite{d2019observation,Guo_2021,Cavicchioli} short-range bosonic mixtures. 
A distinctive property of short-range quantum droplets, which we employ here as a modeling background for the emergence of unprecedented types of 
2D solitary waves, is that they suffer self-evaporation processes~\cite{Ferioli_evaporation,fort2021self}. 
Namely, when excited,  they spontaneously release atoms to attain their lower energy configuration characterized by a fixed intercomponent density ratio. 
The latter, may ensure a reduced single-component extended Gross-Pitaevskii (eGPE) description~\cite{petrov2015quantum,petrov2016ultradilute} 
---see, also the review of~\cite{luo2021new}--- of the genuine two-component setting, which has been experimentally confirmed~\cite{cabrera2018quantum}, and we also use in the present study.

Recently, several properties of droplets have been theoretically investigated. These include ground state phases hosting flat-top~\cite{astrakharchik2018dynamics}, droplet-gas~\cite{Flynn_imbal_2023,Flynn_trap2023,pelayo_two_comp} and multipole~\cite{Kartashov_multipole} configurations, as well as  excitation spectra 
of different types of single~\cite{tylutki2020collective}
or multiple components~\cite{Charalampidis_2comp_drops}. The impact of thermal effects in 1D and three-dimensional (3D) settings~\cite{van2026fragmentation}, as well as the thermodynamics
of such droplets~\cite{mithun2021statistical} have also been studied. 
Additionally, the existence of nonlinear excitations, such as   dark solitons~\cite{katsimiga2023solitary,edmonds2023dark}, shock~\cite{Chandramouli_shocks} and rogue~\cite{chandramouli2025rogue} waves in 
1D settings and vortices~\cite{li2018two,Bougas_vortex_drops,caldara2022vortices,tengstrand2019rotating,yougurt2023vortex} in 2D have been discussed. 
This framework naturally motivates the study of self-trapped structures in the presence of competing attractive and repulsive interactions, where previously unseen regions of stability and dynamics may arise. In addition to standing structures, propagating
ones have been explored both in 1D~\cite{kopycinski2023ultrawide} and in 
higher dimensions~\cite{paredes}. Recently, more exotic settings
have been explored including the behavior of droplets in density-dependent gauge
potentials~\cite{edmonds2025quantumdropletsbeyondmeanfielddensitydependent},
the emergence of spin excitations in  self-bound Bose–Bose droplets~\cite{ritu2026spindensityexcitationsonedimensional}, as well 
as their ability to support topological Hopfion-type 
configurations~\cite{zhao2025stablehopfionstrappedquantum}.

Generally speaking, 2D solitons 
\footnote{Here, we use the term ``soliton'' in a rather loose sense to describe
solitary waves, without implying complete integrability of the pertinent models \cite{ablowitz1991solitons}.}
constitute a fundamentally intriguing, yet highly elusive, class of nonlinear excitations in generic nonlinear dispersive media. Such structures may emerge due to the interplay of dispersion, interaction-induced nonlinearity, and the potential presence of transverse instabilities in higher than 1D~\cite{berge1998wave,frantzeskakis2010dark}. 
Prototypical examples, which we also investigate in the present work and feature markedly different localization properties, are line, lump, and ring solitons, but also dromions~\cite{ablowitz1991solitons,ablowitzrecent}. Specifically, line solitons arising, for instance, within the Kadomtsev–Petviashvili (KP) model~\cite{kadomtsev1970stability} (but also more generally present in
GPE models~\cite{djf2024nonlinear}) are  spatially localized in one direction and extended along a transverse line. Lump solitons are 2D algebraically decaying waveforms that propagate undistorted in integrable systems~\cite{kaup1980inverse,ablowitz1991solitons}. 
In sharp contrast, dromions are 2D exponentially localized configurations, and are solutions of the Davey-Stewartson (DS) equation~\cite{fokas1990dromions,gilson2009dromion}. 
They exist under nontrivial boundary conditions set by a mean flow, and display particle-like interactions~\cite{Fokas_multidimensions,fokas1990dromions}. 
Ring solitons, on the other hand, are well-known to exist
in nonlinear Schr{\"o}dinger (NLS) equation settings~\cite{frantzeskakis2010dark}, and
exist also, e.g., in the cylindrical Korteweg--de Vries (cKdV) equation \cite{johnson1999note}. They can be used to model 
radially symmetric dark solitons in optics \cite{kivshar_dark_1998} and Bose-Einstein condensates (BECs) \cite{kevrekidis2015defocusing}.

Experimental realizations of numerous among the above-mentioned nonlinear configurations remain rather scarce to date in different platforms. For instance, line solitons have been observed in shallow water tanks,  
and their interactions were studied (see, e.g.,  Refs.~\cite{kodama2010kp,yeh2014laboratory}), while the propagation and interaction of lump solitons have only recently been demonstrated in nonlinear optics~\cite{Dieli_lump}. Admittedly, one can argue that earlier studies
observed such structures, see e.g.~\cite{lump2}. 
Dromions, on the other hand, have only been studied in theory~\cite{fokas1990dromions,gilson1991direct}, and 
are still awaiting to be experimentally observed. 

Turning to ultracold atom settings,  which we utilize herein as highly tunable many-body systems, planar dark solitons --being siblings of line solitons-- have been experimentally created in repulsively interacting BECs with the aid of phase-imprinting methods~\cite{denschlag2000generating,burger1999dark}. 
These structures are inherently prone to transverse destabilization yielding their decay into vortex rings~\cite{brian1} --see also 
Refs.~\cite{trombas1,nath2008stability,wen_dark_2013} for relevant stabilization mechanisms. 
Ring solitons, while created experimentally, e.g., in nonlinear optics~\cite{neshev}, are typically found to be unstable and tend to decay into vortex patterns~\cite{kivshar_dark_1998,Theocharis_rings}, while their transient formation has been recently observed in a 2D 
superfluid~\cite{Tamura_rings} causing a 
renewed interest in the subject. 
Interestingly, recent ultracold atom experiments operating in the attractive interaction regime have reported the realization of the celebrated 2D Townes soliton~\cite{Chen_Townes,Bakkali_Townes}. 
This structure is a genuine 2D localized state of the attractive NLS appearing only at specific interaction strengths and exhibiting a scale invariant behavior~\cite{chiao1964self,berge1998wave}.  

Importantly, aside from this specific 2D Townes matter-wave soliton the robust generation of 2D solitons using ultracold atoms is mainly hindered by wave collapse and/or associated instability processes~\cite{berge1998wave,sulem2007nonlinear}. 
This motivates the exploration of alternative platforms -- such as quantum droplets and
their {\it competing} attractive and repulsive interactions-- that may be able to sustain localized 2D soliton configurations. 
Here, it is interesting to emphasize as a supporting argument the recently reported stability of the 1D kink structure in 2D and 3D droplet environments owing to their involved competing repulsive and attractive interactions~\cite{Mistakidis_kinkdrop}.

To address the existence of 2D 
solitons arising in such a setting of competing interactions, we employ the suitable eGPE~\cite{petrov2016ultradilute} with a logarithmic nonlinear interaction term encompassing the LHY and the mean-field nonlinear couplings. 
Recasting the 2D eGPE into its hydrodynamic form, and linearizing the density and phase of the homogeneous droplet background, we identify the chemical potential regions where modulation instability occurs, as well as the ones being stable and, hence, able to support coherent structures. 
In this stable parametric regime, we utilize multiscale analysis which is a central element of our work, to demonstrate the reduction of the eGPE non-integrable droplet model to known integrable nonlinear ones referring to the KP-I, and the DS-I systems. 

Based on these reduced models we identify approximate analytical 2D soliton solutions. Namely, 
line, lump, and ring dark solitons, but also strongly localized dromions that were previously unseen in the droplet background, see also Fig.~\ref{fig2:solutions}. 
Note that some of these states (such as the line soliton) can also be obtained
directly from the 1D solution of the eGPE model (trivially extended along the
transverse direction).
The accuracy of the traveling solutions identified through the above considerations
is testified through direct dynamical simulations of the 2D eGPE showcasing their robustness in the course of the evolution despite the emission of radiation due to their approximate nature.

This work unfolds as follows. 
In Sec.~\ref{sec:theory}, we introduce the appropriate 2D eGPE, along with its hydrodynamic form,  describing symmetric droplets, and analyze the modulational stability of the associated homogeneous background states. 
Based on multiscale expansion methods, in Sec.~\ref{sec:asympt_reductions}, we asymptotically reduce 
the 2D eGPE to effective integrable 
nonlinear models, namely the  
KP-I equation and the DS-I system. 
Section~\ref{sec:sol_der} elaborates on the construction of approximate 2D soliton solutions, i.e., line, lump, and ring solitons, as well as dromions, and showcases their dynamics. 
In Sec.~\ref{sec:conclusions}, we summarize our results and discuss future research directions. 
Finally, Appendix~\ref{app:coeficients} provides details on the coefficients of dromion analytical treatment.

\section{Model setup and linear regime}\label{sec:theory}

\subsection{Droplet background}

We consider a homonuclear bosonic mixture whose components feature intracomponent repulsive 3D scattering lengths $a_{11}^{(3D)}=a_{22}^{(3D)}>0$ and an intercomponent attractive one,  $a_{12}^{(3D)}<0$. 
The system is trapped in a 2D box, of length $L_x=L_y=L$, which can be experimentally achieved using digital-micromirror devices~\cite{Jalm_dynamical_2019,navon2021quantum,Tamura_rings} constraining the motion of the atoms in the $x$-$y$ plane and a tight transverse harmonic trap preventing excitations and pattern formation across the $z$-direction~\cite{Hadzibabic_two_2011}. 
In a corresponding experiment, this composition can be realized, for instance, by the hyperfine states $\ket{1} \equiv \ket{F=1, m_F=-1}$ and $\ket{2} \equiv \ket{F=1, m_F=0}$ of $^{39}$K~\cite{cabrera2018quantum}. 
Accordingly, the individual 2D scattering lengths, $a_{11},a_{22}, a_{12}$, depend on the corresponding 3D ones~\cite{Petrov_interatomic_2001,Bougas_vortex_drops}, which are in turn tunable via Fano-Feshbach resonances~\cite{chin2010feshbach,cheiney2018bright,cabrera2018quantum}. 
Throughout this work, we operate in the scattering length interval where the distance from the mean-field balance point $\delta a \equiv a_{12}+\sqrt{a_{11} a_{22}} \lesssim 0$. 
This refers to the magnetic field region $B \sim 57$G of $^{39}$K~\cite{Errico_Feshbach_2007,Tanzi_Feshbach_2018} in which quantum droplets exist~\cite{luo2021new,petrov2016ultradilute}.

The above condition on the intracomponent scattering lengths ($a_{11}=a_{22}=a$), along with the consideration of equal atom number per component ($N_1=N_2$), suffice to reduce\footnote{Alternatively, this reduction is ensured through $\abs{\psi_{1}}^2/ \abs{\psi_{2}}^2  =  \sqrt{a_{22}}/\sqrt{a_{11}}$~\cite{petrov2016ultradilute,petrov2015quantum,semeghini2018self}.} the two-component system into an effective single-component one~\cite{petrov2016ultradilute} describing a symmetric droplet with 2D wave functions $\psi_1(x,y,t)=\psi_2(x,y,t) \equiv \psi(x,y,t)$. 
The underlying 2D eGPE~\cite{petrov2016ultradilute} reads   
\begin{equation}
i\hbar \psi_t = -\frac{\hbar^2}{2m} \Delta \psi 
+ \frac{8\pi \hbar^2n_0\sqrt{e}}{m\ln^2(a_{12}/a)} |\psi|^2 \psi \ln\big(|\psi|^2\big),
\label{GP}
\end{equation}
and is supplemented with the 
boundary condition
$$\quad |\psi| \rightarrow |\psi_0|={\rm const.}, \quad \text{as} \quad \Vert \mathbf{r}\Vert \rightarrow \infty$$. 
In Eq.~\eqref{GP}, $n_0$ denotes the equilibrium density of the droplet in the thermodynamic limit~\cite{petrov2016ultradilute}, $m$ is the atomic mass, and $\Delta$ represents the 2D Laplacian. 
Below, we employ both Cartesian and cylindrical coordinates, where $\mathbf{r}=(x,y)$ and $\mathbf{r}=(r,\theta)$ respectively, since they will be both used for our analytical considerations in Sec.~\ref{sec:asympt_reductions}, \ref{sec:sol_der}. 
Accordingly, the 2D Laplacian is given by
$\Delta = \partial_x^2+\partial_y^2$, and $\Delta = (1/r)\partial_r(r\partial_r)+(1/r^2)\partial_\theta^2$ respectively.  

Measuring the time $t$, the spatial coordinates $x$ and $y$, and wavefunction $\psi(\mathbf{r},t)$ in units $m/\hbar n_0$, $n_0^{-1/2}$ and $\sqrt{n_0\sqrt{e}}$, respectively, we cast Eq.~\eqref{GP} in the following dimensionless form 
\begin{equation}
i\psi_t + \frac12 \Delta \psi 
- g|\psi|^2 \psi \ln\big(|\psi|^2\big) = 0,
\label{dGP}
\end{equation}
where the effective interaction strength $g=8\pi \sqrt{e}/( \ln^2(a_{12}/a))$~\cite{petrov2016ultradilute,Bougas_vortex_drops}. 
Notice that the logarithmic nonlinearity appearing in the 2D eGPE encapsulates both the standard mean-field interactions and the LHY term. 
Moreover, at low densities the negativity of the logarithm yields attractive interactions, in contrast to large densities where the logarithm is positive and repulsive interactions dominate. 

To numerically study (see below) the evolution of the 2D solitons on top of the droplet background, we utilize the time-dependent 2D eGPE of Eq.~(\ref{dGP}). 
Specifically, a fourth-order Runge-Kutta integrator supplemented by a second-order finite differences method (cross-validated with a pseudospectral method) to compute the spatial derivatives is deployed.  
The spatial and temporal discretizations in our simulations refer to $dx=dy=0.08$ and $dt=10^{-4}$ respectively.
Additionally, homogeneous Neumann boundary conditions suitable for the asymptotics of the 2D line, lump, ring solitons and dromion are used, notably due to the finite density of the associated background. 
Typical box sizes $L=200$ within our dimensionless units translate to $65~\rm{\mu m}$, whilst evolution times of the order of $t \sim 200$ correspond to $14~\rm{ms}$ for transverse oscillator length $a_{z}=0.11~\rm{\mu m}$ with $\omega_{z} = 2\pi \times 20~\rm{kHz}$.

\subsection{Hydrodynamic formulation and stability}
\label{Sec:DH}

\begin{figure}[tbp]
\centering
\includegraphics[width=0.45\textwidth]{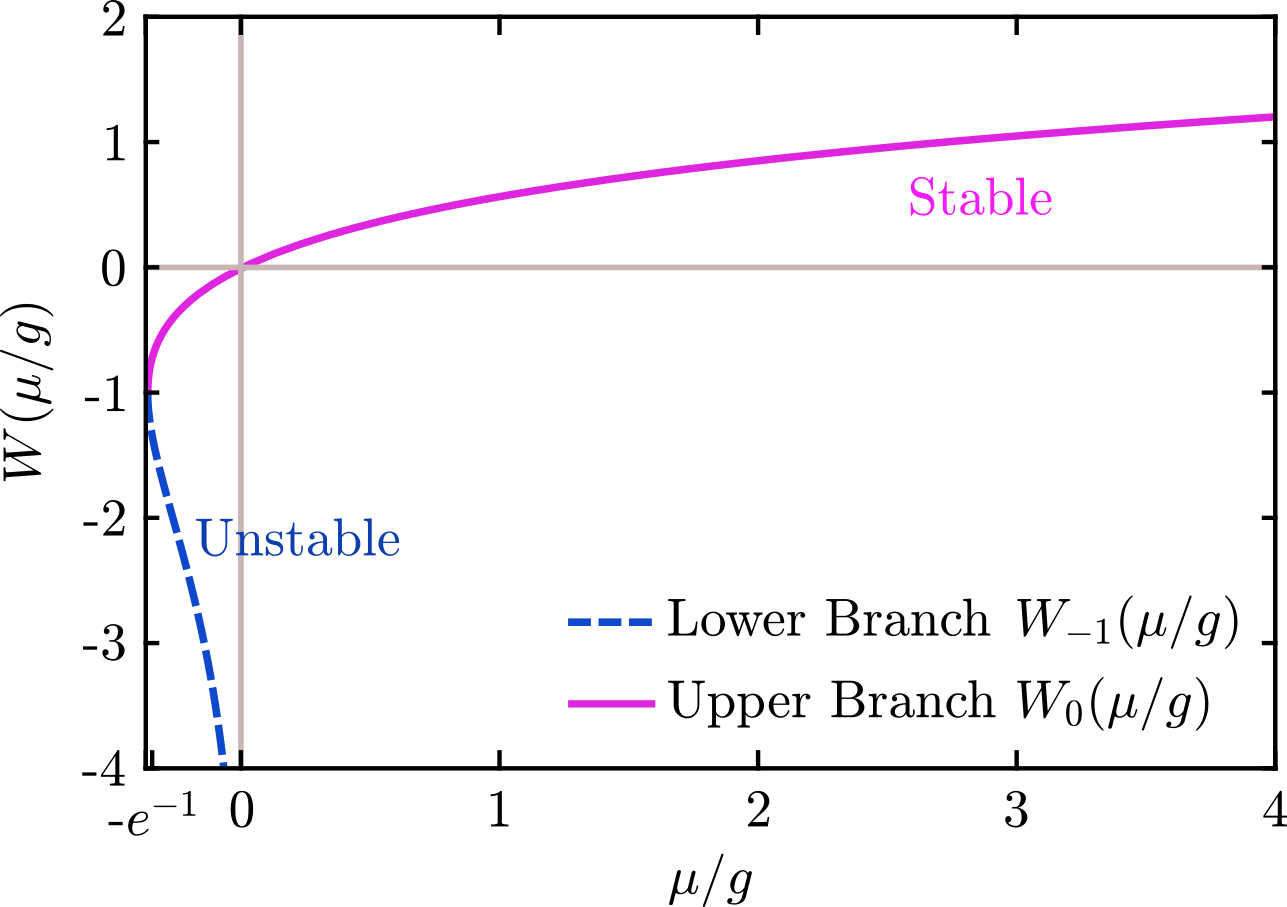}
\caption{The upper and lower  branches of the Lambert $W$ function (see legend) with respect to scaled chemical potential $\mu/g$. 
The upper (lower) branch is associated to modulational stable (unstable) droplet background. The 2D solitons (see also Fig.~\ref{fig2:solutions}) exist on the modulationally stable branch. The horizontal and vertical pink solid lines mark the $W(\mu/g)=0$ and $\mu=0$ cases  respectively.}
\label{Lambert}
\end{figure}

We now introduce the Madelung transformation
\begin{equation}
\psi(\mathbf{r},t) = \sqrt{\rho(\mathbf{r},t)}\, e^{i\phi(\mathbf{r},t)},
\label{Madelung}
\end{equation}
where $\rho(\mathbf{r},t)$ and $\phi(\mathbf{r},t)$ denote the density and phase, respectively. 
Substituting Eq.~\eqref{Madelung} into Eq.~\eqref{dGP}, and separating real and imaginary parts, we obtain the following hydrodynamical system
\begin{subequations}
\begin{eqnarray}
&&\rho_t + \nabla\cdot(\rho\nabla\phi) = 0,
\label{h1}
\\
&&\phi_t + \frac12 \Vert\nabla\phi \Vert^2
- \frac12 \rho^{-1/2}\Delta\rho^{1/2}
+ g\rho\ln\rho = 0.
\label{h2}
\end{eqnarray}
\end{subequations}
The above equations possess the following elementary nontrivial solution, 
$$\rho(\mathbf{r},t)=\rho_0 ={\rm const.}, \quad \phi(\mathbf{r},t)=-\mu t,$$ where $\mu=g\rho_0 \ln (\rho_0)$ is the chemical potential. It can readily be found that the background density $\rho_0$ and the chemical potential $\mu$ are connected through the following equation,
\begin{equation}
\rho_{0}=\frac{\mu/g}{W\!\left(\mu/g\right)},
\label{rho0mu}
\end{equation}
where $W$ is the Lambert $W$-function, namely the inverse function of $f(W)=W\exp(W)$. As shown in  Fig.~\ref{Lambert}, and similarly to the 1D case \cite{katsimiga2023solitary}, there exist two branches: the upper (principal) branch, $W_0(\mu/g)$, and the lower branch $W_{-1}(\mu/g)$, of the Lambert W-function. The upper branch exists for chemical potentials $\mu>\mu_c$, where the critical value $\mu_c=-ge^{-1}$, while the lower branch exists for $-e^{-1}\leq\mu/g\leq 0$.

Next, we consider the stability of the homogeneous background solution. 
Substituting the Ansatz 
\[
\rho = \rho_0 + \varepsilon\rho_1(\mathbf{r},t),
\qquad
\phi = -\mu t + \varepsilon\phi_1(\mathbf{r},t),
\]
into Eqs.~\eqref{h1}-\eqref{h2}, where $0<\varepsilon \ll 1$ is a small parameter, we obtain at $\mathcal{O}(\varepsilon)$ the following linear system for the unknown functions $\rho_1$ and $\phi_1$: 
\begin{subequations}
\begin{eqnarray}
&&\rho_{1t} + \rho_0 \Delta \phi_1 = 0
\label{l1} \\
&&\phi_{1t} - \frac{1}{4\rho_0}\Delta\rho_1
+ g(1+\ln\rho_0)\rho_1 = 0.
\label{l2}
\end{eqnarray}
\end{subequations}

The compatibility condition of Eqs.~\eqref{l1}-\eqref{l2}, $\partial_t \Delta \phi_1 = \Delta \phi_{1t}$, yields
\begin{equation}
\rho_{1tt}
- c^2\,\Delta \rho_1
+ \frac{1}{4}\,\Delta^2 \rho_1
= 0,
\label{B}
\end{equation}
where $c^2=g\rho_0 (1+\ln \rho_0)=\mu\left[1+1/(W(\mu/g))\right]$. 
Obviously, Eq.~(\ref{B}) is either hyperbolic, for $c^2>0$, or elliptic, for $c^2<0$. It can readily be seen that the hyperbolic (elliptic) case corresponds to the upper (lower) branch of the Lambert $W$-function. As a result, plane wave solutions, of the form $\rho_1 \propto \exp[i(\mathbf{k}\cdot \mathbf{r} - \omega t)]$ (where $\mathbf{k}$ and $\omega$ are the wave vector and frequency of the perturbation) are either modulationally stable or unstable, for the upper or the lower branch of $W(\mu/g)$. Indeed, Eq.~(\ref{B}) leads to the Bogoliubov dispersion relation   
\begin{equation}
\omega^2 = c^2\Vert \mathbf{k}\Vert^2
+ \frac14 \Vert\mathbf{k}\Vert^4,
\label{dr}
\end{equation}
with $\omega \in \mathbb{R}~~\forall \Vert k \Vert \in \mathbb{R}$ for $c^2>0$, i.e., for the upper branch, $W_0(\mu/g)$, of the Lambert $W$-function. 
The lower branch is the one that engenders the presence of 2D quantum droplets
that are, arguably, the prototypical structures studied in the context of the
model of Eq.~(\ref{dGP})~\cite{luo2021new}.
Our focus in the present work will shift away from that setting: indeed, we will seek 2D weakly nonlinear excitations (i.e., solitons) on top of the modulationally stable background. 
These will involve structures that asymptote to non-vanishing boundary conditions
(contrary to the vanishing boundary conditions associated with the droplets).

\begin{figure*}
\centering
\includegraphics[width=\linewidth]{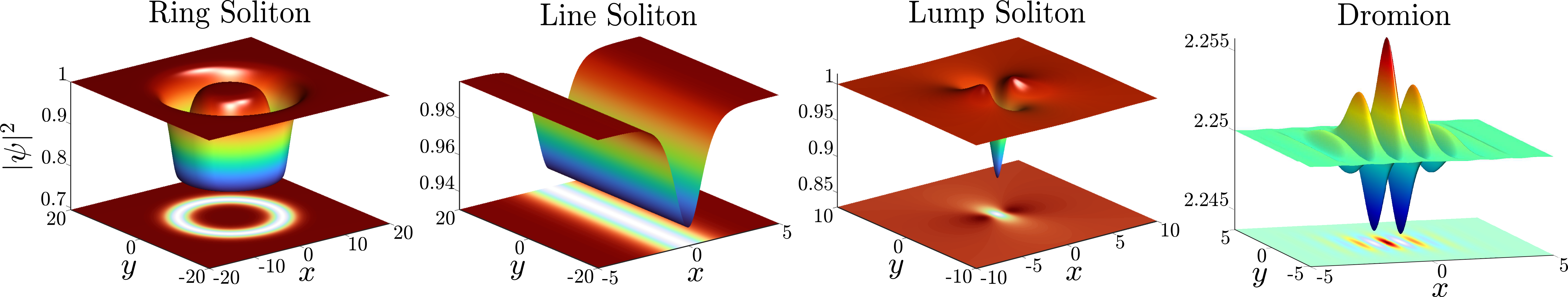}
\caption{Density contours and theoretical projections along the $x$-$y$ plane corresponding to the different solitary wave solutions. The parameters for the ring-soliton are $g=1$, $\xi_0=1$, $\epsilon=0.1$, $\kappa_1=1$, $\rho_0=1$ in Eq.~(\ref{ring}); for the line soliton
the parameters are $g=1$, $\xi_0=0$, $\epsilon=0.1$, $\kappa_1=1$, $\kappa_2=0$, $\rho_0=1$ in Eq.~(\ref{line}). For the lump soliton, we use $g=1$, $\epsilon=0.1$, $\lambda_1=0$, $\lambda_2=1$, $\rho_0=1$ in Eq.~(\ref{lump}), and for the dromion the parameter
values are $g=1$, $\epsilon=0.01$, $k=1.2$, $\rho_0=2.25$, $A=-0.22$, $c_0=-6.48$, $c_1=-1.35$, $c_2=-6.15$, $c_3=-28.64$, $c_4=14.30$ in Eq.~(\ref{dromion}).}
\label{fig2:solutions}
\end{figure*}

\section{Nonlinear regime}\label{sec:asympt_reductions}

Before proceeding with the presentation of the asymptotic reductions of Eq.~\eqref{dGP}, it is useful to briefly comment on the choice of the scales of the independent variables, as well as of the ones pertaining to the unknown fields. All scales are set by a formal  small parameter $0<\varepsilon\ll 1$, which measures the (small)  amplitude of the nonlinear excitation on top of the homogeneous droplet background. The scales of the independent variables are determined according to the long-wavelength approximation of the linear dispersion relation. On the other hand, the scales of the unknown fields are chosen so that maximal balance between dispersion and nonlinearity terms is achieved. This means that leading-order dispersive and nonlinear terms are of the same order (see, e.g. Ref.~\cite{djf2024nonlinear} for a relevant discussion and applications).   

\subsection{Kadomtsev-Petviashvili equations}

Next, we consider the weakly nonlinear regime. First, we show that Eq.~\eqref{dGP} can be asymptotically reduced to a  completely integrable 2D system, namely the KP-I equation. 
We seek solutions of Eqs.~\eqref{h1}-\eqref{h2} in the form of the following asymptotic expansions:
\begin{subequations}
\begin{eqnarray}
\rho &=& \rho_0 + \varepsilon\rho_1 + \varepsilon^2\rho_2 + \cdots,
\label{rhoj_KP} \\
\phi &=& -\mu t + \varepsilon^{1/2}\phi_1
+ \varepsilon^{3/2}\phi_2 + \cdots,
\label{phij_KP}
\end{eqnarray}
\end{subequations}
where the unknown functions $\rho_j$, $\phi_j$ ($j=1,2,\ldots$) depend on the stretched variables 
\[
X = \varepsilon^{1/2}(x-Ct), \quad
Y = \varepsilon y, \quad
T = \varepsilon^{3/2} t,
\]
for the Cartesian case, or
\[
R=\varepsilon^{1/2}(r-Ct),\qquad
\Theta=\varepsilon^{-1/2}\theta,\qquad
T=\varepsilon^{3/2}t,
\]
for the polar coordinates; here, $C$ is an unknown velocity, to be determined in a self-consistent manner. Substituting Eq.~\eqref{rhoj_KP} and \eqref{phij_KP} into Eqs.~\eqref{h1}-\eqref{h2} we obtain
\begin{subequations}
\begin{eqnarray}
&&-c\rho_{j\mathcal{X}}+\rho_0 \phi_{j\mathcal{X}\mathcal{X}} = P_j^{(C,p)},
\label{p1} \\ 
&&-c\phi_{j\mathcal{X}}+g(1+\ln \rho_0)\rho_j =Q_j^{(C,p)},
\label{p2}
\end{eqnarray}
\end{subequations}
where $\mathcal{X}=X$, $P_j^{(C,p)} \equiv P_j^{(C)}$, $Q_j^{(C,p)}\equiv Q_j^{(C)}$ for the Cartesian case, and $\mathcal{X}=R$, $P_j^{(C,
p)} \equiv P_j^{(p)}$, $Q_j^{(C,p)}\equiv Q_j^{(p)}$ for the polar case. For $j=1$, i.e., at $\mathcal{O}(\varepsilon)$, the above system is homogeneous, namely $P_1=Q_1=0$. In this case, combining Eqs.~\eqref{p1}-\eqref{p2} we find 
\begin{equation}
\left[-C^2+g\rho_0(1+\ln \rho_0)\right]\rho_{j\mathcal{X}}=0,  
\label{comp1}
\end{equation}
which, for $\rho_{j\mathcal{X}}\neq 0$, leads to the compatibility condition 
$C^2=g\rho_0(1+\ln \rho_0)$. This result shows that $C\equiv c$, as was obtained in Sec.~\ref{Sec:DH} [see Eq.~\eqref{dr}]. Furthermore, at $\mathcal{O}(\varepsilon)$, we obtain the following equation that connects the functions $\phi_1$ and $\rho_1$:
\begin{equation}
\phi_{1\mathcal{X}}=\frac{c}{\rho_0}\rho_1.
\label{ph1r1}
\end{equation}

For $j>1$, the system of Eqs.~ \eqref{p1}-\eqref{p2} is inhomogeneous, i.e., $P_j^{(C,p)} \neq 0$, $Q_j^{(C,p)} \neq 0$. In this case, combining Eqs.~\eqref{p1}-\eqref{p2} we obtain 
\begin{equation}
\left[-c^2+g\rho_0(1+\ln \rho_0)\right]\rho_{j\mathcal{X}}=cP_j^{(C,p)}+\rho Q_{j\mathcal{X}}^{(C,p)},  
\label{comp2}
\end{equation}
and, thus, the compatibility condition is 
\begin{equation}
cP_j^{(C,p)}+\rho Q_{j\mathcal{X}}^{(C,p)}=0.
\label{comp3}
\end{equation}
First, we consider the Cartesian case. 
For $j=2$, the functionals $P_2^{(C)}$ and $Q_2^{(C)}$ are
\begin{subequations}
\begin{eqnarray}
P_2^{(C)}&=&-\rho_{1T}-\left(\rho_1 \phi_{1X}\right)_X-\rho_0 \phi_{1YY}, 
\label{p1C}\\
Q_2^{(C)}&=&-\phi_{1T}-\frac12 \phi_{1X}^2-\frac{g}{2\rho_0}\rho_1^2+\frac{1}{4\rho_0}\rho_{1XX}.
\label{p2C}
\end{eqnarray}
\end{subequations}
Substituting Eqs.~\eqref{p1C}-\eqref{p2C} into Eq.~\eqref{comp3}, and using Eq.~\eqref{ph1r1}, we obtain the following KP-I  equation:
\begin{equation}
\Big[\rho_{1T}
- \frac{1}{8c}\rho_{1XXX}
+ \frac{g}{2c}(4 + 3\ln\rho_0)\rho_1\rho_{1X}\Big]_{X}
+ \frac{c}{2}\rho_{1YY} = 0.
\label{KP}
\end{equation}
Using the transformations $$T\mapsto -8cT, \quad Y\mapsto c\sqrt{\frac{4}{3}}Y, \quad \rho_1 \mapsto -\frac23g(4+3\ln\rho_0)\rho_1,$$
we express Eq.~\eqref{KP} in the standard form of the KP-I equation, 
\begin{equation}
\left(\rho_{1T}
+ 6\rho_1 \rho_{1X}
+ \rho_{1XXX}\right)_{X}
- 3\rho_{1YY}
=0 .
\label{KP_1}   
\end{equation}

Next, we proceed with the polar case. For $j=2$, the functionals $P_2^{(p)}$ and $Q_2^{(p)}$ are
\begin{subequations}
\begin{eqnarray}
P_2^{(p)}&=&-\rho_{1T}-\left(\rho_1 \phi_{1R}\right)_R-\frac{\rho_0}{cT} \phi_{1R}-\frac{\rho_0}{c^2T^2} \phi_{1\Theta\Theta}, \nonumber \\
\label{p1p}\\
Q_2^{(p)}&=&-\phi_{1T}-\frac12 \phi_{1R}^2-\frac{g}{\rho_0}\rho_1^2+\frac{1}{4\rho_0}\rho_{1RR}.
\label{p2p}
\end{eqnarray}
\end{subequations}
As in the Cartesian case, we substitute Eqs.~\eqref{p1p}-\eqref{p2p} into Eq.~\eqref{comp3}, and employing Eq.~\eqref{ph1r1}, we derive the following, so-called, cylindrical KP-I (cKP-I), or Johnson's \cite{johnson} equation:
\begin{eqnarray}
&&\Big[\rho_{1T}
- \frac{1}{8c}\rho_{1RRR}
+ \frac{g}{2c}(4 + 3\ln\rho_0)\rho_1\rho_{1R}+\frac{1}{2T}\rho_1 \Big]_{R} \nonumber \\ 
&&+ \frac{1}{2cT^2}\rho_{1\Theta\Theta } = 0.
\label{cKP}
\end{eqnarray}
Using the transformations 
\begin{eqnarray}
R\mapsto -\frac{R}{(8c)^{1/3}}, \,\,\, \Theta\mapsto \frac{\Theta}{2\sqrt{3}c^{2/3}}, \,\,\, 
\rho_1 \mapsto -\frac{6c^{2/3}\rho_1}{g(4+3\ln\rho_0)},
\nonumber 
\end{eqnarray}
the cKP-I~\eqref{cKP} is expressed in its standard form,
\begin{equation}
\left(\rho_{1T}
+ 6\rho_1 \rho_{1R}
+ \rho_{1RRR} +\frac{1}{2T}\rho_1\right)_{R}
- \frac{3}{T^2}\rho_{1\Theta\Theta}
=0 .
\label{cKP_1d}   
\end{equation}

\begin{figure*}[ht!]
\centering
\includegraphics[width=1\linewidth]{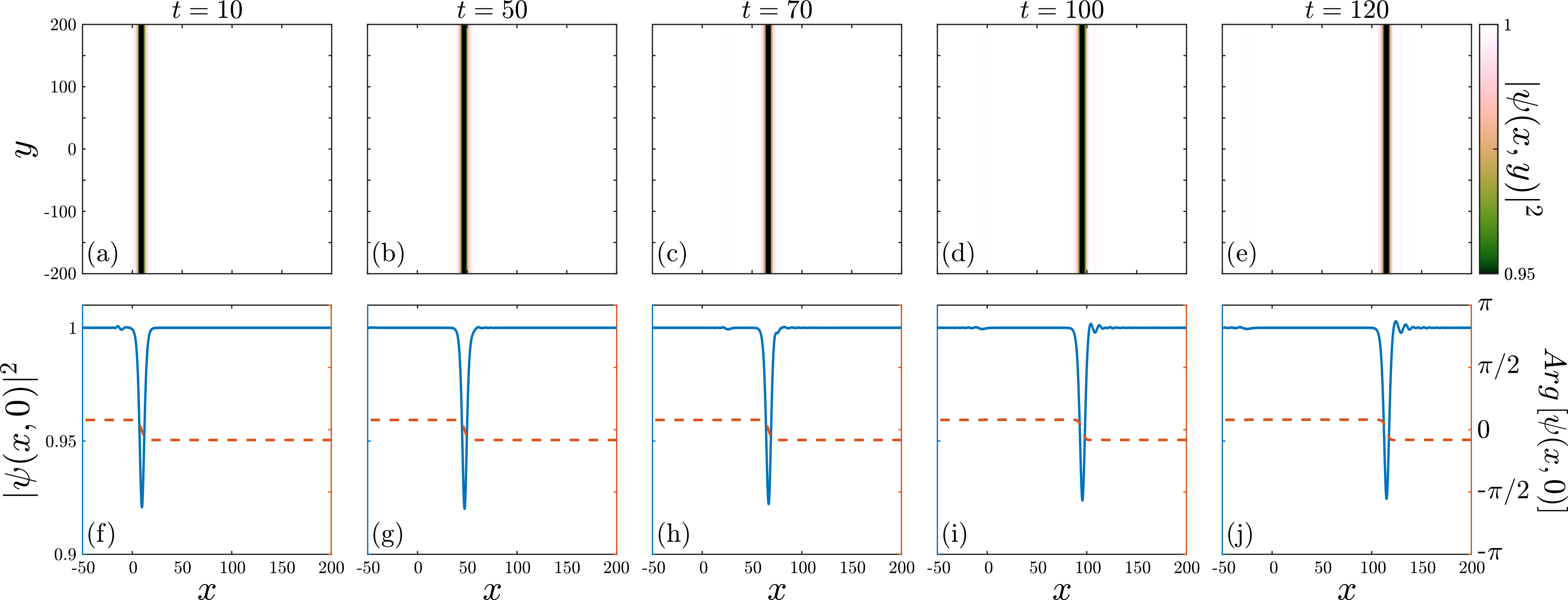}
\caption{Spatiotemporal density evolution of the 2D line soliton described by Eq.~(\ref{line}) at selected time-instants (see legends). 
(a)-(e) 2D density profiles $|\Psi(x,y)|^2$, and (f)-(j) the corresponding 1D density cross-sections along $x$ at $y = 0$ (blue solid line and left axis) together with the associated wavefunction phase $\rm{Arg}[\psi(x,0)]$ (dashed orange lines and right axis). The moving 2D line soliton propagates to the left, while emitting a relatively small amount of radiation. The characteristics of the initial line soliton at $t=0$ are $g=1$, $\xi_0=0$, $\epsilon=0.1$, $\kappa_1=1.2$, $\kappa_2=0$, $\rho_0=1$ respectively.}
\label{fig:line_sol}
\end{figure*}

\subsection{Davey-Stewartson  system}

In the following, we will show that Eq.~\eqref{dGP} can be  asymptotically reduced to another 2D completely integrable system, namely the DS-I system.
To do so, we consider the Cartesian case, and seek solutions of Eqs.~\eqref{h1}-\eqref{h2} in the form of the asymptotic expansions
\begin{subequations}
\begin{eqnarray}
\rho &=& \rho_0 + \varepsilon\rho_1 + \varepsilon^2\rho_2 + \cdots,
\label{rhoj} \\
\phi &=& -\mu t + \varepsilon \phi_1
+ \varepsilon^{2}\phi_2 + \cdots,
\label{phij}
\end{eqnarray}
\end{subequations}
where the unknown functions $\rho_j$, $\phi_j$ ($j=1,2,\ldots$) depend on the stretched variables 
\[
\theta=kx-\omega t, \,
X = \varepsilon^{1/2}(x-Vt),\ 
Y = \varepsilon y, \
T = \varepsilon^{3/2} t,
\]
where $V$ is an unknown velocity, to be determined. Substituting the expansions into Eqs.~\eqref{h1}-\eqref{h2} and equating coefficients of like powers of $\varepsilon$, we obtain the coupled set of equations, 
\begin{subequations}
\begin{eqnarray}
&&-\omega\,\rho_{j\theta}+k^2\rho_0\,\phi_{j\theta\theta}=F_j,\\
&&-\dfrac{k^2}{4\rho_0}\rho_{j\theta\theta}-\omega\,\phi_{j\theta}
+g\rho_j(1+\ln\rho_0)=G_j,
\end{eqnarray}
\end{subequations}
for $j=1,2,\ldots$. 
Notice that, eliminating $\phi_j$, the above system reduces to the following equation for $\rho_j$:
\begin{equation}
-\frac{1}{4}k^2\rho_{j\theta \theta \theta}+\left[g\rho_0(1+\ln\rho_0)k^2-\omega^2\right]\rho_{j\theta}=\omega F_j+\rho_0 k^2 G_{j\theta}. 
\label{sys1}
\end{equation}
Below we focus on the first three orders of approximation, for which the above system leads to the following results.

To leading order, $\mathcal{O}(\varepsilon)$ ($j=1$), the system  is linear and homogeneous 
($F_1=G_1=0$), and yields the dispersion relation for the wavenumber $k$ and frequency $\omega$,
\begin{align}
\omega^2= c^2 k^2 + \frac{1}{4}k^4,
\label{dsr2_DS}
\end{align}
where $c^2=g\rho_0 (1+\ln \rho_0)$ as expected. Note that Eq.~\eqref{dsr2_DS} is identical to Eq.~\eqref{dr} for $k_x=k$ and $k_y=0$. Additionally, at $\mathcal{O}(\varepsilon)$, the leading-order solution for $\rho_1, \phi_1$ reads:
\begin{align}
\begin{cases}
\rho_1=\displaystyle i\frac{k^2\rho_0}{\omega}\,q(X,Y,T)e^{i\theta} +{\rm c.c.},\\[2mm]
\phi_1=q(X,Y,T)e^{i\theta}+ {\rm c.c.} + \Phi(X,Y,T),
\end{cases}
\end{align}
where c.c. denotes complex conjugate. In the above expressions, $q$ denotes the unknown complex envelope of the carrier wave
$\exp(i\theta)$, while $\Phi$ is a long-wavelength mean flow term (corresponding to $k= 0$). Both $q$ and $\Phi$ 
depend solely on the slow variables $X$, $Y$ and $T$ and,  as will be shown below, these unknown functions satisfy
the DS system.

\begin{figure*}[ht!]
\centering
\includegraphics[width=1\linewidth]{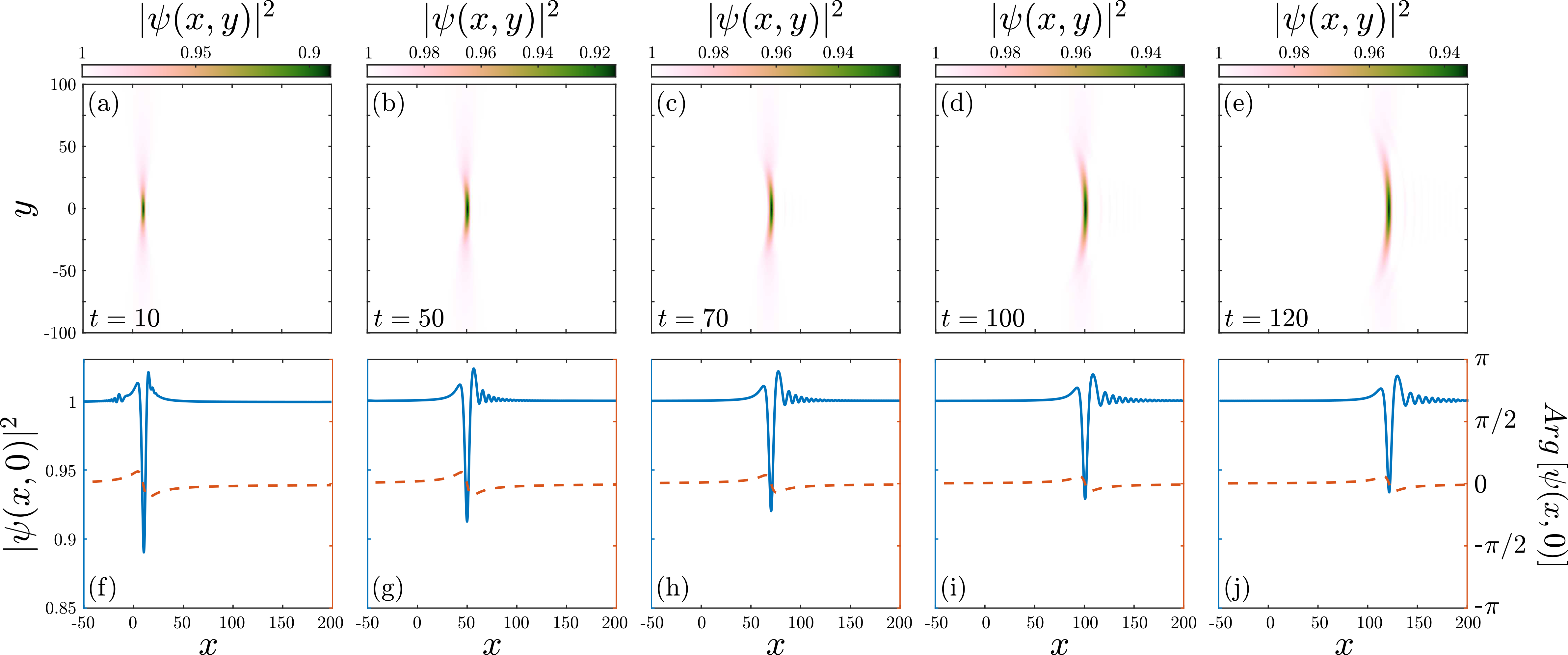}
\caption{(a)-(e) Density snapshots along the $x$-$y$ plane and (f)-(j) their profiles across the $x$-direction at $y=0$  of the 2D lump soliton [Eq.~(\ref{lump})] at different times (see legends) in the course of the evolution. The phase, $\rm{Arg}[\psi(x,0)]$, of the underlying wavefunction is also depicted in  the lower panels with dashed orange lines (with its axis
lying on the right side of the graph). 
As can be seen, the 2D lump moves to the left becoming shallower and acquiring curvature at longer evolution times. 
Emission of radiation occurs simultaneously which is attributed to the approximate nature of the solution in the droplet background.
The parameters of the initial lump configuration correspond to $g=1$, $\epsilon=0.1$, $\lambda_1=0$, $\lambda_2=1$, $\rho_0=1$ respectively.}
\label{fig:lump_sol}
\end{figure*}

Proceeding to the next order $\mathcal{O}(\varepsilon^2)$ ($j=2$), the system becomes inhomogeneous, with
the right-hand side terms given by:
\begin{subequations}
\begin{eqnarray}
F_2&=&c \rho_{1X}-k^2\rho_{1\theta}\phi_{1\theta}-2k\rho_0 \phi_{1\theta X}-k^2\rho_1\phi_{1\theta\theta},
\nonumber \\
G_2&=&-\frac{g}{2\rho_0}\rho_1^2+c\phi_{1X}
-\frac{k^2}{8\rho_0^2}\rho_{1\theta}^2
\nonumber \\ 
&-&\frac{k^2}{2}\phi_{1\theta}^2
+\frac{k}{2\rho_0}\rho_{1\theta X} -\frac{k^2}{4\rho_0^2}\rho_1\rho_{1\theta\theta}. \nonumber
\end{eqnarray}
\end{subequations}
The solvability condition for Eq.~\eqref{sys1} at this order, obtained upon removing the secular terms $\propto \mathrm{exp}(i\theta)$, yields the unknown velocity $V$. The latter is found to be the group velocity $v:= d\omega/dk$, as can be obtained from the dispersion relation 
\eqref{dsr2_DS}. In this way, we find 
\begin{align}
V\equiv v
=\frac{k^3}{4\omega}+\frac{\omega}{k}.
\end{align}
Moreover, at $\mathcal{O}(\varepsilon^2)$, we obtain the following expressions for the unknown fields $\rho_2, \phi_2$:
\begin{eqnarray}
\begin{cases}
\rho_2=\bigl(a_1 q^2e^{2i\theta}+a_2 q_X e^{i\theta}+\text{c.c.}\bigr)+a_3|q|^2+a_4\,\Phi_X,\\
\phi_2=a_5 q^2e^{2i\theta},
\end{cases}
\label{aj}
\end{eqnarray}
where the coefficients $a_j$ ($j=1,2,\ldots 5$) are given in Appendix~\ref{app:coeficients}.

Finally, at the order $\mathcal{O}(\varepsilon^3)$, the right-hand side terms of the system are given by 
\begin{subequations}
\begin{eqnarray}
F_3&=&-\rho_{1T}-\rho_0\phi_{1YY}+c\rho_{2X}-\rho_0\phi_{1XX}  
-k\phi_{1X}\rho_{1\theta} \nonumber \\
&-&k\rho_{1X}\phi_{1\theta}
-k^2\rho_{2\theta}\phi_{1\theta}-k^2\rho_{1\theta}\phi_{2\theta}
-2k\rho_1\phi_{1\theta X} \nonumber \\
&-&2\rho_0k \phi_{2\theta X}-k^2\rho_2\phi_{1\theta\theta}-k^2\rho_1\phi_{2\theta\theta}, \nonumber
\\
G_3&=&\frac{g}{6\rho_0^2}\rho_1^3 -\frac{g}{\rho_0}\rho_1\rho_2-\phi_{1T} \nonumber
+\frac{1}{4\rho_0}\rho_{1YY}+c\phi_{2X}
\\\nonumber
&+&\frac{1}{4\rho_0}\rho_{1XX}-\frac{k}{4\rho_0^2}\rho_{1X}\rho_{1\theta} +\frac{k^2}{4\rho_0^3}\rho_1\rho_{1\theta}^2
-\frac{k^2}{4\rho_0^2}\rho_{1\theta}\rho_{2\theta}
\nonumber \\
&-&k\phi_{1X}\phi_{1\theta} -k^2\phi_{1\theta}\phi_{2\theta} \nonumber 
-\frac{k}{2\rho_0^2}\rho_1\rho_{1\theta X} + \frac{k}{2\rho_0}\rho_{2\theta X}
\\ 
&+&\frac{k^2}{4\rho_0^3}\rho_1^2 \rho_{1\theta\theta}
 - \frac{k^2}{4\rho_0^2}\rho_2\rho_{1\theta\theta}
-\frac{k^2}{4\rho_0^2}\rho_1\rho_{2\theta\theta}. \nonumber
\end{eqnarray}
\end{subequations}
At this order, the solvability condition is derived upon removing again the secular terms, which are now $\propto \exp(i\theta)$ and $\propto \exp(i0\theta)=1$ (constant terms). This way, we obtain the following set of two-coupled equations for the unknown fields $q$ and $\Phi$:
\begin{subequations}
\begin{eqnarray}
&&ic_0 q_T+c_1 q_{XX}+c_2 q_{YY}+\bigl(c_3|q|^2+c_4\Phi_X\bigr)q=0, 
\hspace{1cm}
\label{dr1} \\
&&c_5\Phi_{XX}+c_6\Phi_{YY}+c_7(|q|^2)_{X}=0,
\hspace{1cm}
\label{dr2}
\end{eqnarray}
\end{subequations}
where the coefficients $c_j$ ($j=0,1,\ldots,7$) are functions of $k$. The form of these coefficients, in the long-wavelength limit, are provided in Appendix~\ref{app:coeficients}. 

\begin{figure*}[ht!]
\centering
\includegraphics[width=1\linewidth]{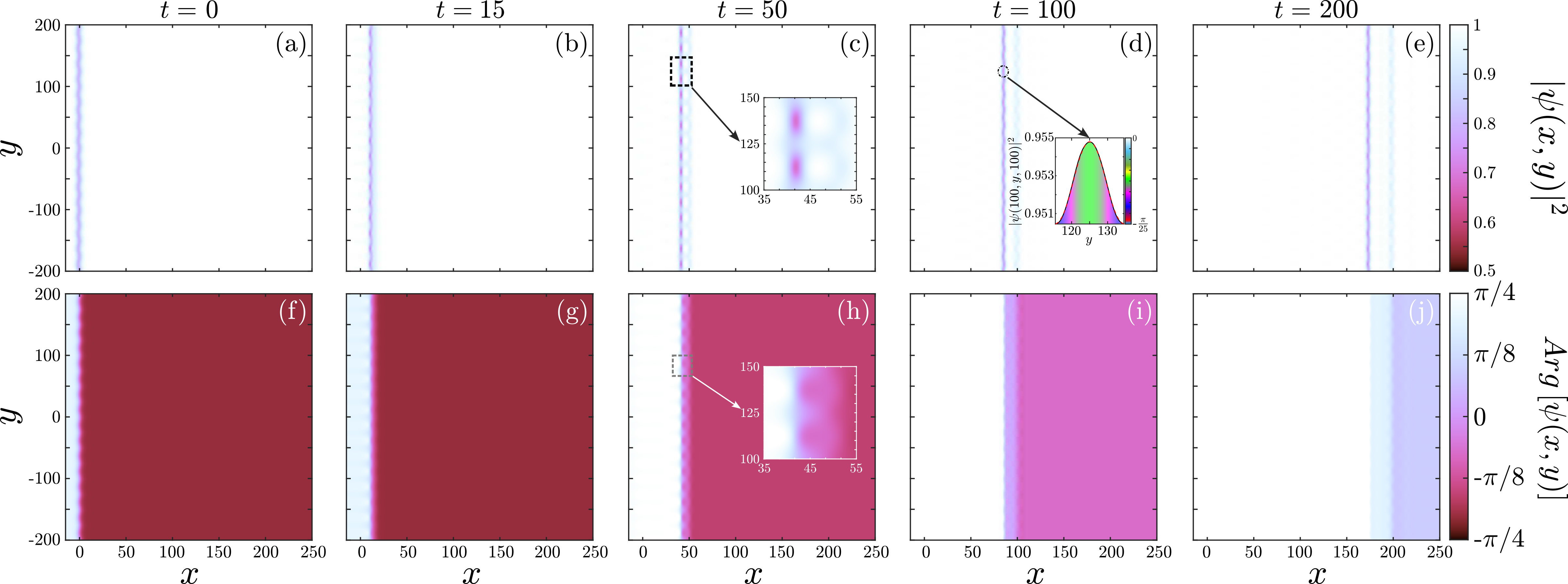}
\caption{(a)-(e) Density and (f)-(j) phase profiles of the 2D bent line soliton at different evolution times (see the individual panels). 
The gradual development of transverse destabilization manifests itself through the dissolvement of the line soliton into an array of lump solitons characterized by phase jumps much smaller than $\pi$. 
The insets in panels (c) and (h) show a magnification of the density and phase, respectively, of two selected lump solitons. The inset in panel (d) presents the density profile of a single lump soliton across $x$ and at $y=0$. The colormap mark the underlying phase of the structure. The parameters of the initial line soliton are the same as in Fig.~\ref{fig:line_sol}, while the bent amplitude corresponds to $A=0.5$ and $K=0.25$ (see text).}
\label{fig:bent_line}
\end{figure*}

The system of Eqs.~\eqref{dr1}-\eqref{dr2} has been used to describe modulated multidimensional water waves in finite depth. This model was first derived by Benney and Roskes (BR)~\cite{BR}, then by Davey and Stewartson (DS)~\cite{DS}, and also by Djordjevic and Redekopp~(DR)~\cite{DR}, who included surface tension. In its general form the BR/DR system is not integrable. Nevertheless, in the shallow water limit ($kH \rightarrow 0$, where $k$ is the wavenumber and $H$ is the water's depth), Ablowitz and Segur~\cite{AblowitzSegur1979} derived the completely integrable version, known as the DS system.

To derive the nearly integrable version of Eqs.~\eqref{dr1}-\eqref{dr2} we proceed as follows. First, we introduce the transformations 
\begin{eqnarray}
&&X\mapsto\sqrt{\left|\frac{c_0}{c_1}\right|}\,X,
\qquad~~
Y\mapsto\sqrt{\left|\frac{c_0}{c_2}\right|}\,Y,
\nonumber \\
&&|q|^2\mapsto 2A\left|\frac{c_0}{c_3}\right|\,|q|^2, 
\quad 
\varphi\mapsto-\frac{c_0}{c_4}\,\varphi,
\nonumber 
\end{eqnarray}
where the parameter $A$ is given by
$$A=-\frac{2+\ln \rho_0}{8+6\ln \rho_0},$$
and then express Eqs.~\eqref{dr1}-\eqref{dr2} in the following normalized form:
\begin{subequations}
\begin{eqnarray}
&&i q_{T}+ q_{XX}+ q_{YY}-|q|^2 q-\varphi q = R[q],
\label{nds1}\\
&&\varphi_{XX}-\varphi_{YY}+2(|q|^2)_{XX}=0,
\label{nds2}
\end{eqnarray}
\end{subequations}
where $\varphi=\Phi_X$ and the functional perturbation $R[q]$ is given by
$$R[q]=-\left(1+2 A\right)|q|^2 q.$$ 
In the case $R[q]=0$, the system of Eqs.~~\eqref{nds1}-\eqref{nds2} takes the form of the completely integrable DS-I system; note that in DS-I, Eq.~\eqref{nds1} [Eq.~\eqref{nds2}] is elliptic (hyperbolic) as concerns the 2nd derivative operator in $q$ ($\varphi$). 
In such a case, the DS-I system supports exact analytical 2D soliton solutions, known as dromions, which will be presented in Sec.~\ref{sec:dromion}. Such solutions are still physically relevant in the case where  the functional $R[q]$ may be considered as a small perturbation for values of the background density $\rho_0 \gtrsim e^{-1}$ or chemical potential $\mu \gtrsim- ge^{-1}$, i.e., close to the critical value $\mu_c$. 
This in turn implies that the dromion for any $\mu > - ge^{-1}$ is an approximate solution and, as such, it is expected that it will deviate the most from the numerical predictions; see also the discussion in Sec.~\ref{sec:dromion} and Fig.~\ref{fig:dromion}.

\section{Soliton solutions}\label{sec:sol_der}

\subsection{Line and lump solitons}

Starting with the KP-I soliton solutions in the Cartesian case, there exist two physically important types of soliton solutions, namely line and lump solitons.
The single line soliton of the KP-I equation, expressed in the original coordinates $x$, $y$, $t$, reads
\begin{subequations}
\begin{eqnarray}
\rho_1&=&-\frac{3\kappa_1^2}{g(4+3\ln\rho_0)}{\rm sech}^2\xi, 
\label{line} \\ 
\xi&=&\varepsilon^{1/2}\kappa_1 \Bigg[x-c\left(1-\varepsilon \frac{4\kappa_1^2-3\kappa_2^2}{8c^2} \right)t
\nonumber \\
&+&\varepsilon^{1/2}\frac{\sqrt{3}\kappa_2}{2c}y \Bigg]-\xi_0,
\end{eqnarray}
\end{subequations}
where $\kappa_1$ and $\kappa_2$ are arbitrary $\mathcal{O}(1)$ parameters, and $\xi_0$ sets the initial soliton location. 
We remark that line soliton solutions or their  traveling gray soliton variants~\cite{katsimiga2023solitary}  are also expected to exist  in the corresponding effective 1D eGPE model featuring logarithmic nonlinearity~\cite{Mistakidis_kinkdrop}. This approach would lift one level of approximation and it is an interesting future research direction.

\begin{figure*}[ht!]
\centering
\includegraphics[width=1\linewidth]{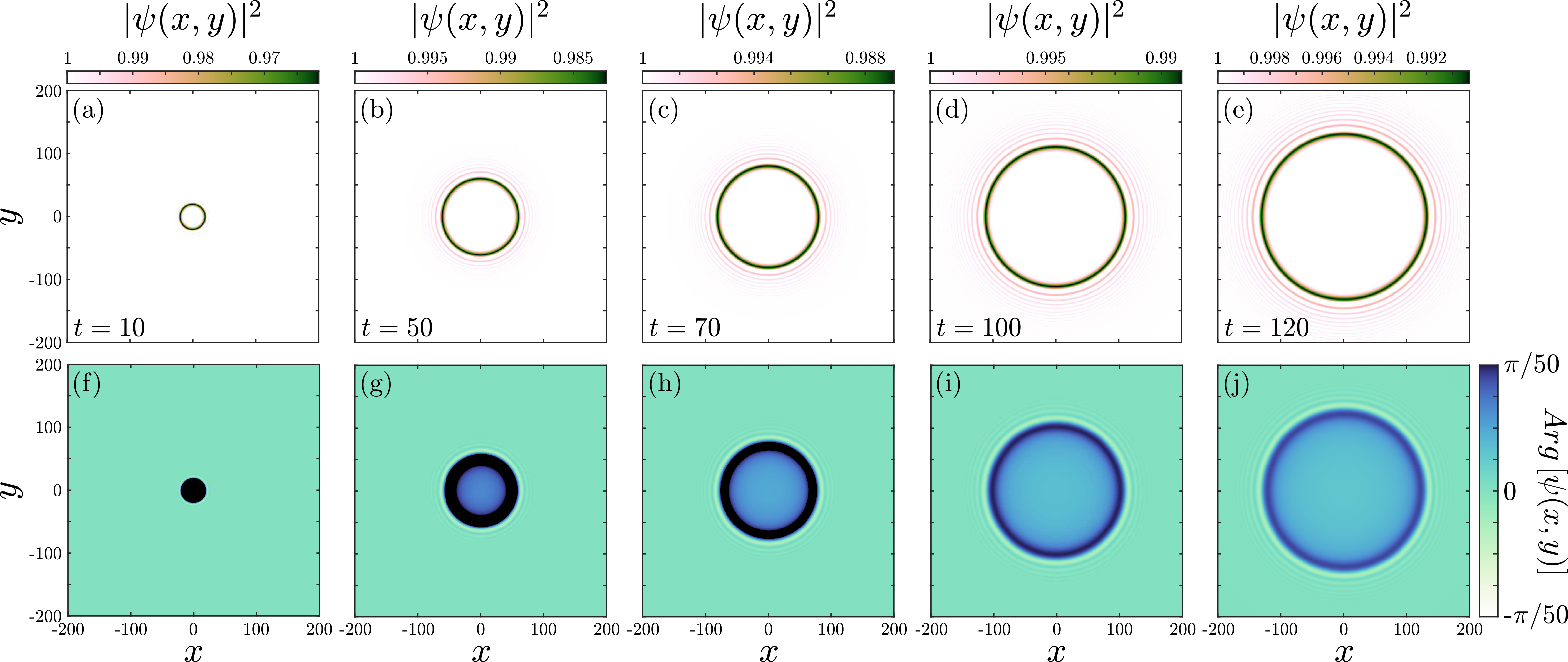}
\caption{(a)-(e) Density profiles across the $x$-$y$ plane of the 2D ring soliton at selected times (see legends). 
(f)-(j) The phase of the 2D wavefunction, $\rm{Arg}[\psi(x,y)]$ showing a phase jump across the radius of the soliton that is gradually decreasing during the evolution; 
the ring expands outwards, becoming shallower and emitting radiation in the form of radially symmetric excitations in a form reminiscent of a radial dispersive shock wave~\cite{Hoefer_DSW}. 
The characteristics of the initial ring soliton refer to $g=1$, $\xi_0=1$, $\epsilon=0.1$, $\kappa_1=0.6$, $\rho_0=1$, respectively.}
\label{fig:ring_sol}
\end{figure*}

On the other hand, the KP-I equation possesses weakly localized 2D soliton solutions, decaying  algebraically at infinity. These solutions are known as lumps~\cite{ablowitz1991solitons};
indeed, the line solitons of KP-I are transversely unstable toward the formation of
the lumps. 
The one-lump soliton solution, expressed in the original variables $x,y,t$, is given by the following rational expression:
\begin{equation}
\!\!\!\!
\rho_1
=
-\frac{6}{g\,(4+3\ln\rho_0)}
\frac{-(x'+\lambda_1 y')^2+\lambda_2^2 y'^2+\frac{1}{\lambda_2^2}}
{\left[(x'+\lambda_1 y')^2+\lambda_2^2 y'^2+\frac{1}{\lambda_2^2}\right]^2},
\label{lump}
\end{equation}
where $\lambda_1$ and $\lambda_2$ are arbitrary $\mathcal{O}(1)$ parameters, and the variables $x'$ and $y'$ are given by:
\begin{subequations}
\begin{eqnarray}
x'&=&\varepsilon^{1/2}\left\{
x-\left[c-\varepsilon\frac{3}{8c}(\lambda_1^2+\lambda_2^2)\right]t
\right\},
\\
y'&=&
\varepsilon \frac{\sqrt3}{2c}\left(y-\varepsilon^{1/2} \frac{\sqrt{3}}{2}\lambda_1 t
\right).
\end{eqnarray}
\end{subequations}
The above expressions for the line and the lump solitons can be used to construct pertinent approximate soliton solutions ---valid up to $\mathcal{O}(\varepsilon)$--- of the original problem of Eq.~\eqref{dGP},
\begin{eqnarray}
\!\!\!\!\!
\psi &\approx&[\rho_0+\varepsilon \rho_1(x,y,t)]^{1/2}
\nonumber \\
&\times&\exp\left[-{\rm i}\mu t +{\rm i} \varepsilon^{1/2}\frac{c}{\rho_0}\int^X \rho_1(X',y,t) {\rm d}X'\right].
\label{KPlump} 
\end{eqnarray}

The density evolution of a relatively shallow 2D line soliton, as obtained from the approximate KP-I solution of Eq.~(\ref{line}), on top of the droplet background is depicted in Fig.~\ref{fig:line_sol}. 
Inspecting the 2D density profiles in Fig.~\ref{fig:line_sol}(a)-(e), it becomes apparent  that the structure is a dark (grey) soliton stripe, of the form of a density depletion localized along the propagation $x$-direction, and being elongated transversely. 
It is also characterized by a phase jump $\rm{Arg}[\psi(x,0)] \approx \pi/4$ [Fig.~\ref{fig:line_sol}(f)]  across its core, which is smaller than $\pi$ (that corresponds to a stationary (black) soliton) due to its shallow depth. 
The soliton moves to the right, preserving its shape, while emitting a small amount of  radiation (subsequently propagating on the homogeneous background) traced back to the approximate nature of the solution. 
The aforementioned radiation occurs in both right and left $x$-directions, while it clearly manifests as small amplitude oscillatory patterns on the droplet background identified in the corresponding density profiles across the $x$ axis at $y=0$ illustrated in Fig.~\ref{fig:line_sol}(f)-(j). 
Additionally, the depth and phase jump of the moving line soliton reduce slightly in the course of the evolution, see Fig.~\ref{fig:line_sol}(f)-(j). 
This confirms the validity of the asymptotic reduction to the KP-I model in the modulational stable droplet environment. 
Nonetheless, the line soliton structure is persistent over long times without suffering significant distortions, thereby suggesting its dynamical stability and, hence, the feasibility to observe this solitary wave in 2D droplet systems.

Turning to the dynamics of the 2D lump soliton described by Eq.~(\ref{lump}), we observe a somewhat more involved response [Fig.~\ref{fig:lump_sol}] as compared to the previously discussed straight line soliton. 
Indeed, as time evolves the lump which is a strongly localized density depression with algebraic spatial decay moves to the left [Fig.~\ref{fig:lump_sol}(a)-(e)], while experiencing a slight gradual broadening of its core and becoming noticeably less deep, see in particular the 1D density profiles in Fig.~\ref{fig:lump_sol}(f)-(j).
This process, being more prominent when compared to the line soliton propagation, is naturally accompanied by a reduction of the initial small phase jump across its core, i.e., $\rm{Arg}[\psi(x,0)] \approx \pi/4$, due to the shallow character of the structure. 
In sharp contrast to the line soliton, the initially symmetric localized lump soliton profile develops a slight curvature, indicating the deformation of its shape during the evolution. 
This behavior, along with the simultaneous development of small amplitude radiation is a result of the approximate character of the analytical solution in the 2D eGPE model. 
Summarizing, the robustness of the localized lump configuration as time progresses validates the relevance of the KP-I lump solution for describing 2D solitary waves in quantum droplet environments, while the observed deformations highlight the important role of non-integrable effects in long-time dynamics.

An interesting connection between the line and the lump solitons can also be established through the transverse destabilization of the former, as it was also  discussed in polariton superfluids~\cite{Frantzeskakis_lump}. 
To demonstrate this behavior in the current droplet setting we slightly modify the core of the line soliton by perturbing it along the $y$-direction through 
$x \to x- A\cos(Ky)$ 
where $A=0.5$ and $K=0.25$. 
The ensuing density dynamics of the perturbed line soliton within the eGPE framework is shown in Fig.~\ref{fig:bent_line}(a)-(e) along with the associated phase profiles in Fig.~\ref{fig:bent_line}(f)-(j). 
It can be readily seen that as time evolves the line soliton moves to the left with the azimuthal perturbation becoming more pronounced. 

As a result, noticeable spatial deformations develop on top of the line waveform eventually leading to its dissolvement into an array of lump solitons [Fig.~\ref{fig:bent_line}(c)-(e)] with a characteristic phase structure shown in Fig.~\ref{fig:bent_line}(h)-(j). 
The resulting lump solitons subsequently propagate robustly for long evolution times. 
For better visualization of their spatial structure, we provide in the inset of Fig.~\ref{fig:bent_line}(c), (h) the 2D density distribution and phase of two selected lumps participating in the array and located in the interval $(x,y) \in [35, 55]\times[100, 150]$. 
Their structure is in line with the analytical lump soliton configuration as shown in the inset of Fig.~\ref{fig:bent_line}(d) which provides a magnification of the density along the $y \in [100, 125]$ region at $x=100$ illustrating a single lump configuration of the array together with its phase represented by the colormap. Here, fitting the analytical lump soliton solution to the numerical waveform reveals good agreement; this further supports the fact that the destabilization products are lump solitons and confirms the relevance of the KP-I lump solution.

\begin{figure*}[ht!]
\centering\includegraphics[width=1\linewidth]{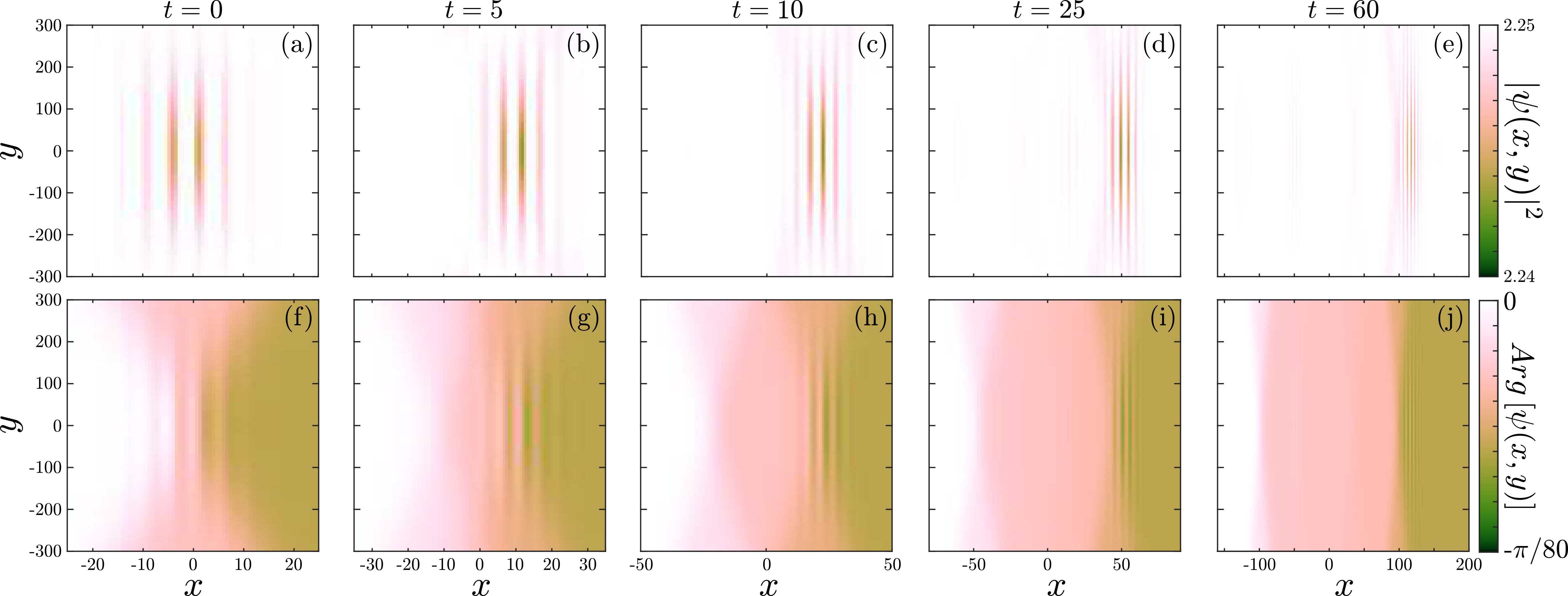}
\caption{(a)-(e) Density snapshots across the $x$-$y$ plane of the dromion configuration at specific evolution times (see legends) along with (f)-(j) the corresponding phase profiles. 
As  can be seen, the initial five hump dromion waveform moves to the right maintaining its overall envelope structure, while emitting radiation traveling to the left. 
A relative shallow phase jump occurs across its hump of the dromion. 
Here, we use $g=1$ for the eGPE, while for the initial dromion solution we use $\epsilon=0.01$, as well as $k=1.2$, $\rho_0=2.25$; for these values  we obtain $c_0=-6.48$, $c_1=-1.35$, $c_2=-6.15$, $c_3=-28.64$, $c_4=14.30$, and $A=-0.22$. Furthermore, we use $\lambda_r=1,\ \mu_r=1,\ \lambda_i=0,\ \mu_i=0,\ \nu=1,\ \xi_0=0,\ \eta_0=0,\ \theta_0=0$. }
\label{fig:dromion}
\end{figure*}

\subsection{Ring dark solitons}
We now proceed with the radially symmetric solution stemming from the cKP-I \eqref{cKP}, which is in fact a soliton solution of the cKdV equation. The latter supports an exact ring-shaped soliton solution (usually called ``cylindrical soliton''), which can be found by the formal reduction of the cKdV to the usual 1D KdV model~\cite{hirota1979exact}. The relevant solution is a ring-shaped pulse of a $\text{sech}^2$ profile, on top of a rational background $\propto R / T$. Obviously, the divergence of the background at initial times (i.e., for $T \rightarrow 0$, see, e.g.,~\cite{hirota1979exact,mannan2015ring}) makes the above exact solution difficult to investigate either numerically or experimentally. Nevertheless, a much more convenient form of the cylindrical soliton is available. Indeed, an asymptotic analysis, valid for solutions of sufficiently large radii~\cite{johnson1999note,KO}, shows that an approximate form of the cylindrical soliton's core can be well approximated by the planar KdV soliton, but with a slowly varying amplitude due to the expansion of the solution. This approximate form of the soliton in terms of the original variables $r$ and $t$ is as follows:
\begin{subequations}
\begin{eqnarray}
\rho_1&=&-\frac{3\eta^2}{g(4+3\ln\rho_0)}
{\rm sech}^2\xi,
\label{ring} \\[1mm]
\xi&=&\varepsilon^{1/2}\eta
\Bigg[r-c\left(1-\varepsilon \frac{3\eta^2}{2c^2}\right)t\Bigg]-\xi_0,
\label{ring_xi}
\end{eqnarray}
\end{subequations}
where $\eta^2=\kappa_1^2\left(\frac{t_0}{t}\right)^{2/3}$ corresponds to the slowly varying amplitude of the cylindrical soliton, with  $\kappa_1$ being an arbitrary $\mathcal{O}(1)$ parameter, and $\xi_0$ being a free parameter corresponding  to the center of the ring.

As a next step, we examine the dynamics of the 
ring dark soliton when embedded in the homogeneous background of the eGPE model;
see Fig.~\ref{fig:ring_sol}. 
This radially symmetric nonlinear excitation is characterized by a circular symmetric density depression with a nearly uniform core chosen here to be relatively shallow to corroborate its moving nature. 
During the evolution, the ring structure preserves its radial shape but expands outward increasing its core whose amplitude reduces simultaneously [Fig.~\ref{fig:ring_sol}(a)-(f)]. 
Accordingly, the initial phase jump across the ring core becomes smaller over time, see Fig.~\ref{fig:ring_sol}(f)-(j).
The ring expansion demonstrates the important role of the curvature in this configuration (in contrast to the above-discussed line and lump solitons), while emission of weak radiation in the form of outward propagating circular wavefronts 
is observed. 
Overall, we infer that in-spite of the aforementioned noticeable shape deformations of the ring in the course of the evolution, it is still an adequately persistent structure in droplet media maintaining its phase pattern. 
This justifies the underlying asymptotic cKP-I reduction to capture the essential response of the ring soliton.

\subsection{Dromions}\label{sec:dromion}

We finally consider the approximate dromion solution of Eq.~\eqref{dGP}, stemming from the DS-I system of Eqs.~\eqref{nds1}-\eqref{nds1} for $R[q]=0$. This solution reads: 
\begin{eqnarray}
\psi
&=&\Bigg\{\rho_0-\varepsilon\Bigg[
\frac{k^2\rho_0}{\omega}\sqrt{\frac{2|c_0 A|}{|c_3|}}\, \nonumber
q\,e^{i(kx-\omega t)}+\text{c.c.}
\Bigg]\Bigg\}^{1/2}
\\ \nonumber
&\times&
\exp\Bigg\{i\Bigg[
-\mu t
+\varepsilon\Bigg(i
\sqrt{\frac{2|c_0 A|}{|c_3|}}\,
 q\,e^{i(kx-\omega t)} \\
&+&\text{c.c.}-\frac{c_0}{c_4}\sqrt{\frac{|c_1|}{|c_0|}}
\int \varphi(X,Y,T)\,dX
\Bigg)\Bigg]\Bigg\}.\label{dromion}
\end{eqnarray}
In the above expressions, the field $q$ is given by
\begin{align}
q(X,Y,T)=\frac{Q_1(X,Y,T)}{Q_2(X,Y,T)},
\end{align}
where 
\begin{subequations}
\begin{eqnarray}
Q_1(X,Y,T)&=&\nu\sqrt{\lambda_r \mu_r}\;
\exp\left(-i(\mu_i \hat{\xi}+\lambda_i \hat{\eta})\right)
\nonumber \\
&\times&
\exp\left(\frac{i}{2}(|\mu|^2+|\lambda|^2)T- i\theta_0 \right),
\\
Q_2(X,Y,T)&=&\cosh\!\big(\mu_r(\hat{\xi}-\xi_0)\big)\cosh\!\big(\lambda_r(\hat{\eta}-\eta_0)\big) 
\nonumber \\
&+&\frac{|\nu|^2}{4}\exp\!\big(\mu_r(\hat{\xi}-\xi_0)\big)\exp\!\big(\lambda_r(\hat{\eta}-\eta_0)\big). \nonumber \\
\end{eqnarray}
\end{subequations}
Here, the coordinates $\hat{\xi}$ and $\hat{\eta}$ are given by
\begin{eqnarray}
\hat{\xi}&=&\xi-2\lambda_iT, \quad \xi= \frac{1}{\sqrt{2}}(X+Y), 
\nonumber \\
\hat{\eta}&=&\eta -2\lambda_i T,
\quad \eta = \frac{1}{\sqrt{2}}(X-Y),
\end{eqnarray}
while the constants $\xi_0$ and $\eta_0$ are 
$$\xi_0=-\frac{1}{\mu_r}\ln\!\Big(\frac{|\mu|}{\sqrt{2\mu_r}}\Big), \quad 
\eta_0=\frac{1}{\lambda_r}\ln\!\Big(\frac{|\lambda|}{\sqrt{2\lambda_r}}\Big),$$
with $\lambda=\lambda_r+i\lambda_i \in \mathbb{C}$, $\mu=\mu_r+i\mu_i \in \mathbb{C}$ (with $\lambda_r,\mu_r \in \mathbb{R}^+$), $\theta_0 \in \mathbb{R}$ and $\nu \in \mathbb{C}$ being constant parameters.

In the $\phi$ component, the dromion solution is characterized by the boundary conditions
\begin{subequations}
\begin{eqnarray}
\varphi_1&=&\lim_{\xi \rightarrow -\infty} \varphi = 2\lambda_r^2 {\rm sech}^2(\lambda_r(\hat{\eta}-\eta_0)),
\label{bc1} \\
\varphi_2&=&\lim_{\eta \rightarrow -\infty} \varphi = 2\mu_r^2 {\rm sech}^2(\mu_r(\hat{\xi}-\xi_0)),
\label{bc2}
\end{eqnarray}
\end{subequations}
showing that the dromion $q$ is located at the intersection of two line solitons in the $\varphi$ component. Based on these boundary conditions, the mean flow $\varphi$ is expressed in terms of the density of the surface mode $|q|^2$ as
\begin{eqnarray}
\varphi(X,Y,T)=&-&(|q|^2 +\varphi_1+\varphi_2) 
\nonumber \\
&+&\int_{-\infty}^\xi |q|^2_\eta d\xi'+ \int_{-\infty}^\eta |q|^2_\xi d\eta'.    
\end{eqnarray}

The spatiotemporal evolution of the above dromion solution embedded in the 2D eGPE background is shown in Fig.~\ref{fig:dromion}. 
Monitoring the density snapshots [Fig.~\ref{fig:dromion}(a)–(e)] it becomes evident that the original multi-hump (here consisting of five peaks, see also Fig.~\ref{fig2:solutions}) dromion configuration propagates to the right, while largely preserving its overall localized structure and envelope. 
This response confirms the robustness of the waveform as an approximate solution of the 2D eGPE model. 
Indeed, the approximate nature of the solution manifests itself through the fact that there is a simultaneous emission of radiation in the opposite direction, a behavior that is traced back to  the non-integrable property of the eGPE framework (as well as the non-exactness of the
dromion solution). 
The associated phase profiles are illustrated in Fig.~\ref{fig:dromion}(f)-(j) revealing the existence of relatively shallow phase variations across the involved humps which is consistent with the chosen low amplitude character of the dromion. 
Summarizing, we can infer that dromions remain strongly localized and dynamically persistent waveforms in the 2D droplet background, while featuring deformations and  radiation which, however, do not destroy their coherent composition.

\section{Conclusions and Perspectives}\label{sec:conclusions}

We have investigated the existence of 2D 
solitons embedded in a homogeneous background in an eGPE model that
can support quantum droplets. Here, we extend studies to unprecedented (to the best of our knowledge) configurations
in the self-repulsive regime, characterized by large background densities. 
The atomic states of interest are chosen to be symmetric, i.e., the density ratio of the  ensuing two-components assumes a fixed value proportional to the intracomponent interactions, and hence it can be described by a single-component eGPE with logarithmic nonlinearity encompassing the mean-field interaction and the LHY contribution. 
To appreciate the parametric regions of possible existence of coherent structures, we first 
derive the hydrodynamic form of the 2D eGPE, and subsequently perform linearization in terms of the density and the phase to classify the stability of the homogeneous droplet background with respect to the chemical potential. 

Focusing on the underlying modulationally stable region (which represents
the regime of dominant self-repulsion), we systematically deploy a multiscale analysis aiming to reduce the original non-integrable 2D eGPE model to effective integrable ones. 
It is demonstrated that the associated asymptotic descriptions correspond to the Kadomtsev-Petviashvili-I and Davey-Stewartson-I equations that capture the long-wavelength dynamics of the system. 
These reduced effective models enable us to analytically extract approximate 2D soliton solutions. 
The latter include dark line and dark lump solitons, radially symmetric ring dark solitons, and envelope dromions, which feature distinct localization behavior ranging from algebraic (lump) to exponential (dromion) decay. 
The validity of the extracted solutions is assessed through direct dynamical simulations based on the original 2D eGPE, showing that the waveforms propagate robustly for long evolution times despite the emission of weak radiation emanating from their approximate nature. 

Particularly, for all nonlinear structures considered herein we find that the deviation between our analytical solutions and the time evolved states is of the order of $\varepsilon^2$ at the initial stages of the dynamics. These aberrations  gradually grow, as expected, for longer times. 
Specifically, line, lump and ring solitons remain in closest agreement with the theoretical predictions, although they progressively deviate due to the emission of 
dispersive radiation, whereas dromions exhibit the largest departure from their analytical waveform despite approximately preserving their overall shape.
Nevertheless, the states appear to be dynamically robust and long-lived allowing their experimental observability. 

Our results can inspire a multitude of promising forthcoming investigations on  multidimensional 
solitons in eGPE systems and, more broadly, nonlinear models 
featuring competing repulsive and attractive interactions such as the cubic-quintic
NLS~\cite{michinel1}. 
A natural first step is to appreciate the spectral stability of the identified configurations, in particular ring dark solitons~\cite{Theocharis_rings,Tamura_rings}, and the existence of their underlying solution branches in the presence and absence of an external trap.  
The latter may not only impact the parametric regions of existence of the respective structures, as it was shown, for instance, for dark solitons in droplets~\cite{katsimiga2023solitary}, but importantly it can facilitate their stability and even act against transverse destabilization~\cite{lee2026numericalidentificationstationarystates}. 
Along these lines, understanding the manifestation of transverse instabilities especially, e.g., for line or lump solitons through computing the Bogoliubov-de Gennes excitation spectrum but also via constructing the appropriate variational framework which has been done for dark soliton stripes in droplets~\cite{Bougas_vortex_drops} and repulsive Bose gases~\cite{Ma_DSS,Feder_dark_2000} is desirable. This can lead to enriched higher-dimensional soliton dynamics and the generation of topological defects.

In the context of fully localized states, it is equally interesting to explore the dynamics of multihump dromion configurations~\cite{hietarinta1990multidromion} in competing interaction systems.  
Another highly intriguing direction pertains to the study of interactions among the 2D solitary waves uncovered herein, including line-line, lump-lump, dromion-dromion and mixed line-lump structures whose description may require the development of suitable effective particle models that have been demonstrated in repulsive gases~\cite{Kevrekidis_DSS,kamchatnov2011condition}. 
Additionally, extensions to 3D settings offer a promising avenue for unveiling unprecedented multidimensional soliton states,  
such as spherical shell solitons~\cite{Wang_spherical} and vortex tangles~\cite{Middleton_tangles} which have been investigated in repulsive Bose gases but are yet to be unraveled in competing interaction models.

\begin{acknowledgements}
S.I.M acknowledges support by the Army Research Office (ARO) under Award number: W911NF-26-1-A043. 
P.G.K was supported by the U.S. National Science Foundation under the award PHY-2408988 and a grant from the Simons Foundation SFI-MPS-SFM-
00011048.

\end{acknowledgements}

\appendix

\section{Coefficients of the DS equation}\label{app:coeficients}

The coefficients $a_j$ ($j=1,2,\ldots,5$) appearing in Eq.~\eqref{aj} are given by:
\begin{subequations}
\begin{align}
&a_1=-\frac{-3\rho_0 k^8+4g\rho_0^{2}k^6+4\rho_0 k^4\omega^2}
{6\omega^2k^4},
\\
&a_2=\frac{-\rho_0k^5+4\rho_0k \omega^2}{4\omega^3},
\\
&a_3=\frac{k^6-4g\rho_0k^4-4k^2\omega^2}{4g(1+\ln\rho_0)\omega^2},
\\
&a_4=\frac{k^4+4\omega^2}{4g(1+\ln\rho_0)k\omega},
\\
&a_5=-\,i\,\frac{\omega}{2\rho_0k^2}a_1.
\end{align}
\end{subequations} 

The coefficients $c_j$ ($j=0,1,\ldots,7$) in Eqs.~\eqref{dr1}-\eqref{dr2} are approximated in the long-wavelength limit (i.e., keeping terms corresponding to the smaller power of $k$ for each $c_j$). Furthermore assuming, without loss of generality, the case of right-going waves with $c>0$, the coefficients $c_j$ are given by:

\begin{subequations}
\begin{align}
&c_0=-2\rho_0k^2,
\\
&c_1=-c_5=-\frac{3\rho_0 }{4c}k^3,
\\
&c_2=c_6 = -\rho_0 ck,
\\
&c_3=-\frac{2g^2 \rho_0^3}{3c^3}\left(8+10\ln\rho_0+3\ln^2\rho_0\right)k^3
,\\
&c_4=-c_7= \frac{\rho_0(4+3\ln\rho_0)}{1+\ln\rho_0}k^3.
\end{align}
\end{subequations}


\bibliographystyle{apsrev4-1}
\bibliography{reference}

\end{document}